\documentstyle[12pt,epsf]{ioplppt}
\begin{document}
%
%
%
%
%
%
\jl{1}
\title{Extrapolation procedure for low-temperature series
for the square lattice spin-1 Ising model}[Extrapolation of 
spin-1 Ising series]

\author{I Jensen\ftnote{1}{E-mail address: iwan@maths.mu.oz.au} and
A J Guttmann\ftnote{2}{E-mail address: tonyg@maths.mu.oz.au}}

\address{Department of Mathematics,
The University of Melbourne, Parkville, Victoria 3052, Australia.}

\begin{abstract}
The finite-lattice method of series expansions has been combined
with a new extrapolation procedure to extend the low-temperature
series for the specific heat, spontaneous magnetisation and 
susceptibility of the spin-1 Ising model on the square lattice.
The extended series were derived by directly calculating the series 
to order 99 (in the variable $u=\exp[-J/k_B T])$ and using the new 
extrapolation procedure to calculate an additional 13--14 terms. 
\end{abstract}

\pacs{05.50.+q, 64.60.Fr, 02.70.-c}

\maketitle

\section{Introduction}

In a recent paper \cite{EGJ} we reported on the calculation
and analysis of low-temperature
series for the square lattice spin-1 Ising model. The 
series was derived to 79th order (in the variable $u=\exp[-J/k_B T])$
using the finite-lattice method 
\cite{DNE} and employed a new algorithm which removed much of the 
memory-size restrictions of previous implementations. In this paper we 
report on a further extension of these series to order 113 for
the specific heat and spontaneous magnetisation and order 112
for the susceptibility. The extension is obtained by direct 
calculation of the series to order 99 and use of a new extrapolation 
procedure to extend the series by an additional 13 or 14 terms. 
The improvement in the direct series derivation is due to 
a more efficient implementation of the algorithm and the use of
parallel computation.
The extrapolation procedure is similar to and inspired
by work on directed percolation \cite{BGDP}. 

\section{The series expansion technique}

The Hamiltonian defining the spin-1 Ising model in a 
magnetic field $h$ can be written:
\begin{equation}
{\cal H} = J\sum_{\langle ij \rangle}(1-\sigma_i \sigma_j)+
  h\sum_{i}(1-\sigma_i)
	\label{hamiltonian}
\end{equation}
\noindent 
where the spin variable $\sigma_i = 0,\; \pm 1$.
The first sum is over nearest 
neighbour pairs and the second sum is over sites. The constants
are chosen so the ground state ($\sigma_i = +1 \; \forall \; i$)
has zero energy. 
The low temperature expansion is based on perturbations from the 
fully aligned ground state.
The expansion is expressed in terms of the temperature variable
$u = \exp(-\beta J)$ and the field variable 
$\mu = \exp(-\beta h)$, where $\beta = 1/k_B T$.
The expansion 
of the partition function in powers of $u$ may be expressed as
\begin{equation}
Z = \sum_{n=0}^{\infty} u^{n}\Psi_{n}(\mu)
\end{equation}
\noindent 
where $\Psi_{n}(\mu)$ are polynomials in $\mu$. It is more
convenient to express the field dependence in terms of the
variable $x=1-\mu$ and truncate the expansion at
$x^{2}$
\begin{equation}
Z = Z_{0}(u) + xZ_{1}(u) + x^{2}Z_{2}(u) + \ldots,
\end{equation}
where $Z_n(u)$ is a series in $u$ formed by collecting all
terms in the expansion of $Z$ containing factors of $x^n$.
Standard definitions yield the spontaneous magnetisation 

\begin{equation}
M(u) = M(0) + \frac{1}{\beta} \left. 
 \frac{\partial \ln Z}{\partial h}\right|_{h=0}
  = 1 + Z_{1}(u)/Z_{0}(u),
\end{equation}
since $x=0$ in zero field. For the zero-field susceptibility we find
\begin{equation}
\fl
\chi(u) = \left. \frac{\partial M}{\partial h}\right|_{h=0} =
\frac{\partial}{\partial h}\left. \left(
\frac{1}{\beta Z}\frac{\partial Z}{\partial h}\right)\right|_{h=0} =
\beta \left[2\frac{Z_{2}(u)}{Z_{0}(u)}-
\frac{Z_{1}(u)}{Z_{0}(u)}-
\left( \frac{Z_{1}(u)}{Z_{0}(u)} \right)^{2}\right].
\end{equation}

The specific heat series is derived from the zero field partition
function (via the internal energy
$U= -(\partial/\partial \beta) \ln Z_{0}$),

\begin{equation}
C_{v}(u) = \frac{\partial U}{\partial T} =
\beta^{2}\frac{\partial^{2}}{\partial \beta^{2}}\ln Z_{0}=
(\beta J)^{2}\left( u\frac{\mbox{d}}{\mbox{d}u} \right)^{2}
\ln Z_{0}(u).
\end{equation}

So in order to obtain the series expansion of the specific heat,
spontaneous magnetisation and susceptibility it suffices to
calculate the three quantities $Z_0$, $Z_1$ and $Z_2$. 

On the square lattice
the infinite lattice partition function $Z$ can be 
approximated by a product of partition functions $Z_{mn}$ on 
{\em finite} ($m\! \times \! n$) lattices,
\begin{equation}
Z(u) \approx \prod_{m,n} Z_{mn}(u)^{a_{mn}} 
\mbox{\hspace{6mm} with } m \leq n \mbox{ and } m+n \leq r,
\end{equation}
\noindent
where $r$ is a cut-off which limits the size of the rectangles
considered. The weights $a_{mn}$ are small integers and are known 
explicitly \cite{E78} for the square lattice, 

\begin{equation}
a_{mn} = \left\{ \begin{array}{rl} 1 & \hbox{if $m+n = r$} \\
 -3 & \hbox{if $m+n = r-1$} \\
 3 &  \hbox{if $m+n = r-2$} \\
 -1 & \hbox{if $m+n = r-3$} \\
  0 & \hbox{otherwise} \end{array} \right.
\end{equation}
Due to the symmetry of the square lattice one obviously has
that $Z_{mn} = Z_{nm}$ so one need only consider the case
$m \leq n$ and change the weights $a_{mn}$ appropriately, i.e.,
multiply by 2 if $m < n$.

For the low-temperature expansion of the 
Ising model $Z_{mn}$ is
calculated as the sum over all spin 
configurations on the finite lattice. All spins 
outside the $m \! \times \! n$ range are fixed at $+1$.
The number of terms derived correctly 
with the finite lattice method is given by the power of the 
lowest-order connected graph not contained in any of the rectangles 
considered, which in this case
are chains of sites all in the `0' state. From
the Ising Hamiltonian we see that such chains give rise to 
terms of order $3r+1$.
For a given value of $r$ the series expansion is thus correct 
to order $3r$. 

The efficient way of calculating $Z_{mn}$ is by transfer matrix 
techniques. We refer to \cite{EGJ} for a detailed description
of the algorithm. For this work we used a more efficient
implementation of the algorithm and a parallel
computer and  were able to derive the series directly up to
a maximal cut-off $r_m = 33$.

\section{Extrapolation of series}

The series can be extended significantly
via an extrapolation method similar to that of \cite{BGDP}.
Consider the series for $Z_n (u)$. For each $r \leq r_m$ 
we use the finite-lattice method to calculate the polynomials
$Z_{n,r}(u) = \sum_{j=0} z_{n,j,r}u^j$ correct to ${\cal O}(u^{3r+20})$.
As already noted these polynomials agree with the series for $Z_n$
to ${\cal O}(u^{3r})$. Next, we look at the sequences $d_{n,r,s}$ obtained
from the difference between successive polynomials
\begin{equation}
\fl
Z_{n,r+1}(u)-Z_{n,r}(u) = u^{3r+1} \sum_{s \geq 0} 
(z_{n,3r+s+1,r+1}-z_{n,3r+s+1,r})u^s
= u^{3r+1} \sum_{s \geq 0} d_{n,s,r} u^s.
\end{equation}
\noindent
The first of these correction terms $d_{n,0,r}$ is often a
simple sequence which one can readily identify. In the case
of $Z_0$ we find the sequence
\[
-d_{0,0,r} = 1,2,6,18,52,138,338,778,1712,\ldots
\]
from which we conjecture
\begin{equation}
  d_{0,0,r} = -2^{r+2}+(r^3+3r^2+2r+18)/3, \;\; r \geq 2.
\end{equation}
\noindent
The formula for $d_{0,0,r}$ holds
for all the $r_m \! - \!1$ values that we calculated and we are
very confident that it is correct for all values of $r$. As was the
case in \cite{BGDP} the higher-order correction terms $d_{n,s,r}$
can be expressed as rational functions of $d_{n,0,r}$. Due to
the form of the first correction term this leads to the general
extrapolation formulae 
\begin{equation}\label{eq:ef}
 d_{n,s,r} = \frac{1}{6s!n!}\sum_{j=0}^{s+n+3}a_{n,s,j}r^j
+\frac{1}{s!n!}\sum_{j=0}^{s+n}b_{n,s,j}r^j2^r, \;\; r \geq s+2.
\end{equation}
\noindent
The factors in front of the sums have
been chosen so as to make the leading coefficients particularly
simple. We were able to find formulae for
all correction terms up to $s=13$ for $Z_0$ and $Z_1$ and 
up to $s=12$ for $Z_2$. The coefficients in the extrapolation
formulae are listed in tables \ref{Z0corr}--\ref{Z2corr}.

\begin{table}
\caption{\label{Z0corr} Coefficients $a_{0,s,j}$ and $b_{0,s,j}$ in the
extrapolation formula \protect{\eref{eq:ef}} for $Z_0$.}
\epsfxsize=19cm \epsffile[50 600 600 800]{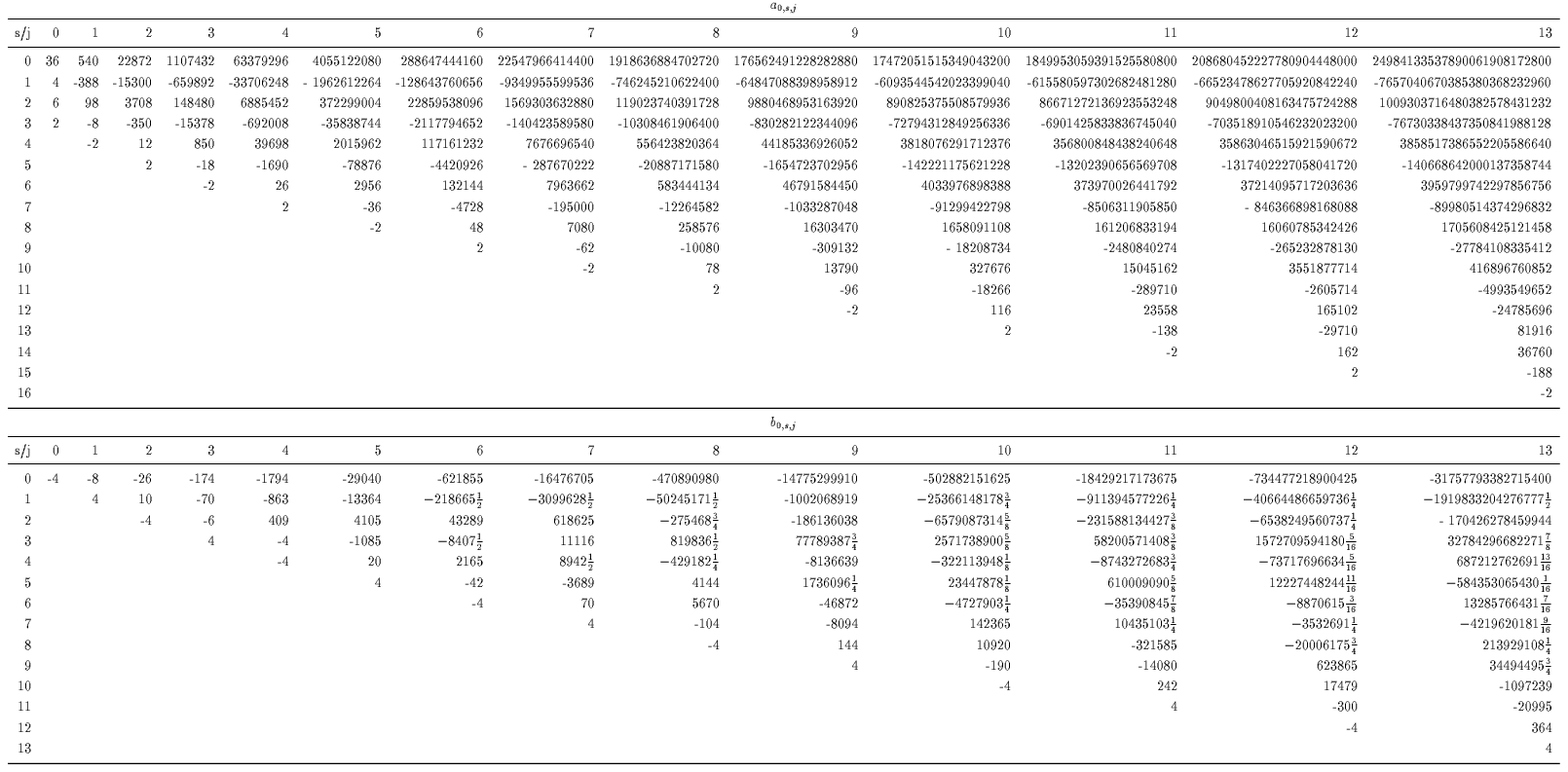}
\end{table}

\begin{table}
\caption{\label{Z1corr} Coefficients $a_{1,s,j}$ and $b_{1,s,j}$ in the
extrapolation formula \protect{\eref{eq:ef}} for $Z_1$.}
\epsfxsize=19cm \epsffile[50 560 600 800]{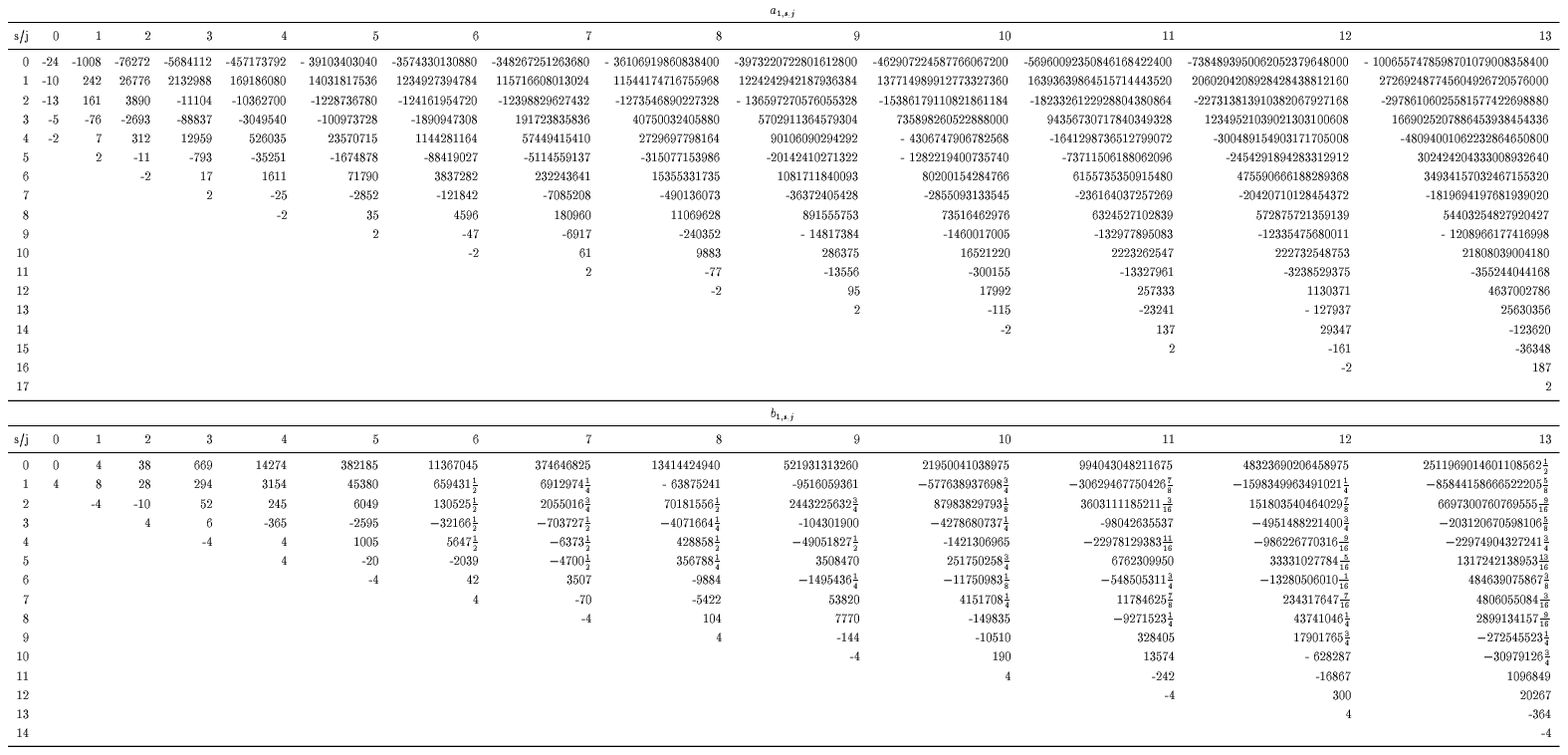}
\end{table}

\begin{table}
\caption{\label{Z2corr} Coefficients $a_{2,s,j}$ and $b_{2,s,j}$ in the
extrapolation formula \protect{\eref{eq:ef}} for $Z_2$.}
\epsfxsize=19cm \epsffile[50 560 600 800]{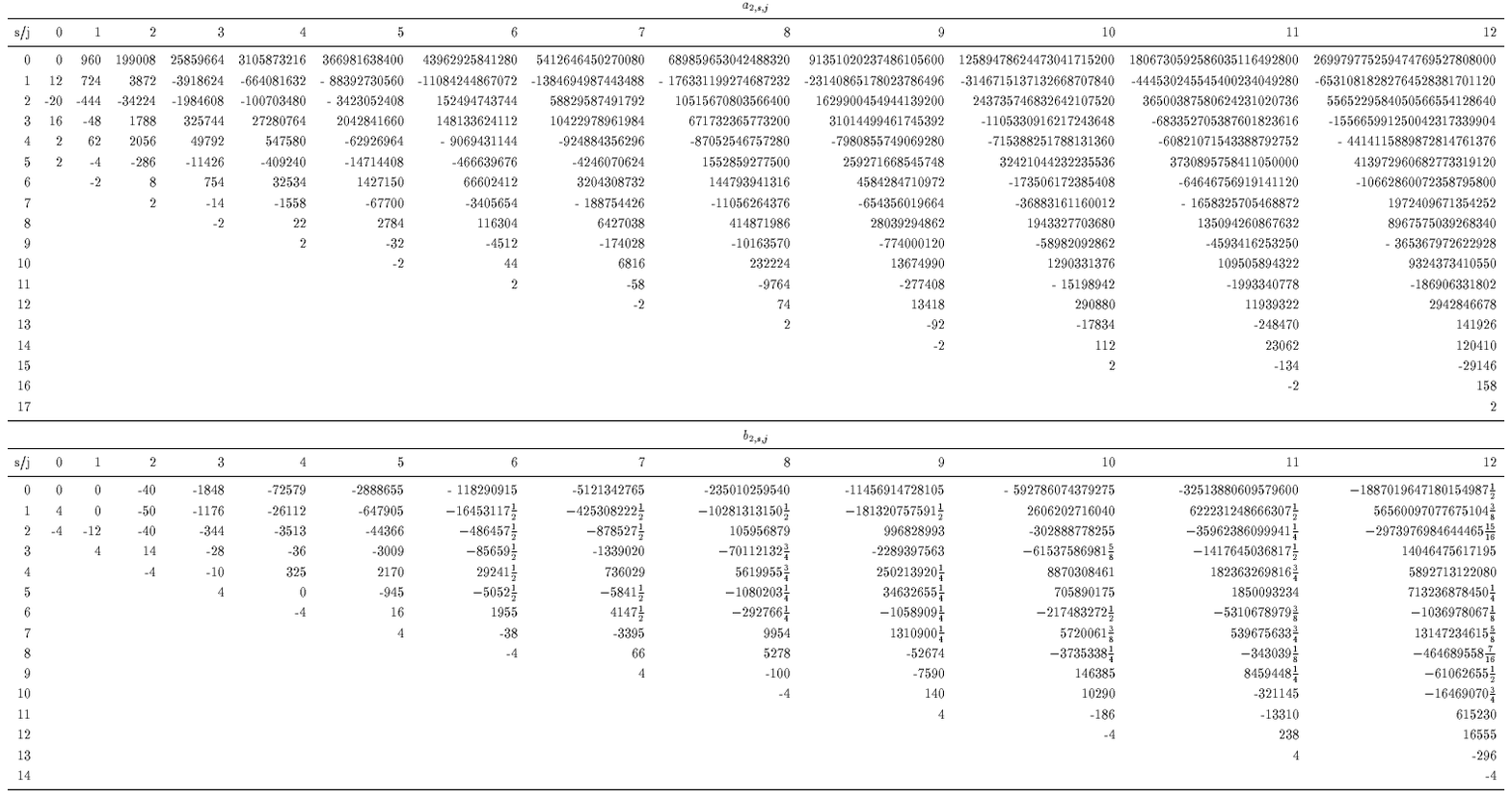}
\end{table}

It is clear from equation \eref{eq:ef} that the $r_m-s-2$
terms available from the various sequences for the correction
terms are not sufficient to determine all the $2(s+n+2)+1$
unknown coefficients in the extrapolation formulae for large $s$. 
However, from tables \ref{Z0corr}--\ref{Z2corr} we immediately see
that the leading coeffients, $a_{n,s,s+n+3}$ and $b_{n,s,s+n}$,
in the extrapolation formulae are alternating in sign 
but otherwise constant, e.g., $a_{n,s,s+n+3} = (-1)^{s+n}2$ and
$b_{n,s,s+n} = -(-1)^{s+n}4$. In general we have found that 
the leading coeffients are expressible as polynomials in $s$,

\begin{equation}\label{eq:ap}
a_{n,s,s+n+3-k} = \frac{(-1)^{s+n}}{(2s)!}
  \sum_{j=0}^{2k+l}\alpha_{n,k,j}s^j,
\end{equation}
and
\begin{equation}\label{eq:bp}
b_{n,s,s+n-k} = \frac{(-1)^{s+n}}{(s+1)!4^{s-1}}
   \sum_{j=0}^{2k}\beta_{n,k,j}s^j, 
\end{equation}
where $l=\max (k-3,0)$. The coefficients of these polynomials
are listed in \tref{Atab} and \tref{Btab}.

\begin{table}
\caption{\label{Atab} The coefficients $\alpha_{n,k,j}$ in the
extrapolation formulae \protect{\eref{eq:ap}}.}
\begin{indented}
\scriptsize\rm
\item[]\begin{tabular}{@{}lrrrrrr}
\br
 k/j& 0 & 1 & 2 & 3 & 4 & 5 \\
\mr
\centre{7}{$\alpha_{0,k,j}$} \\
\mr
0 & 2 & 12 & 96 & 25920 & 0 & 0 \\
1 &  & 2 & -1468 & 510840 & 1300549152 & 6977845103040\\
2 &  & 2 & -558 & -888576 & -2703028424 & -15961597601040 \\
3 & &  & -428 & 600150 & 1803808536 & 12990760770024 \\
4 & &  & 6 & 38850 & -409368974 & -4669644914724 \\
5 & & &  & -7950 & -18927216 & 667412568090 \\
6 & & &  & 126 & 5412484 & 519181770 \\
7 & & & &  & -224616 & -5708659068 \\
8 & & & &  & 6354 & 428756328 \\
9 & & & & & -96 & -15678630 \\
10 & & & & &  & 484770 \\
11 & & & & & & -10656 \\
12 & & & & & & 96 \\
\br
\centre{7}{$\alpha_{1,k,j}$} \\
\mr
0 & 2 & 10 & 312 & 7200 & 967680 & 0 \\
1 & & 2 & -1192 & 566700 & 1554817152 & 8201223021600 \\
2 & & 2 & -522 & -1082526 & -3148467656 & -18452430302760 \\
3 & & & -428 & 592890 & 2028534576 & 14630011358544 \\
4 & & & 6 & 39480 & -411219494 & -5057308922574 \\
5 & & & & -7950 & -20088936 & 678397444650 \\
6 & & & & 126 & 5432476 & 1825419990 \\
7 & & & &  & -224616 & -5791418388 \\
8 & & & &  & 6354 & 430798878 \\
9 & & & & & -96 & -15704550 \\
10 & & & & &  & 484770 \\
11 & & & & & & -10656 \\
12 & & & & & & 96 \\
\br
\centre{7}{$\alpha_{2,k,j}$} \\
\mr
0 & 2 & 4 & 384 & -14400 & 483840 & 0 \\
1 &  & 2 & -940 & 679560 & 1809266592 & 9448276380480 \\
2 & & 2 & -510 & -1261176 & -3582329672 & -20938915839600 \\
3 & &  & -428 & 598470 & 2221265016 & 16212179478024 \\
4 & & & 6 & 39930 & -415205294 & -5412395679564 \\
5 & & &  & -7950 & -20805456 & 691017125850 \\
6 & & & & 126 & 5445412 & 2648223930 \\
7 & & & &  & -224616 & -5853962268 \\
8 & & & & & 6354 & 432269568 \\
9 & & & & & -96 & -15721830 \\
10 & & & & & & 484770 \\
11 & & & & & & -10656 \\
12 & & & & & & 96 \\
\br
\end{tabular}
\end{indented}
\end{table}

\begin{table}
\caption{\label{Btab} The coefficients $\beta_{n,k,j}$ in the
extrapolation formulae \protect{\eref{eq:bp}}.}
\tiny\rm
\begin{tabular}{
@{\hspace{0.2mm}}r
@{\hspace{2mm}}r
@{\hspace{2mm}}r
@{\hspace{2mm}}r
@{\hspace{2mm}}r
@{\hspace{2mm}}r
@{\hspace{2mm}}r
@{\hspace{2mm}}r
@{\hspace{2mm}}r
@{\hspace{2mm}}r}
\br
 k/j& 0 & 1 & 2 & 3 & 4 & 5 & 6 & 7 & 8\\
\mr
\centre{10}{$\beta_{0,k,j}$} \\
\mr
0 & -1 & 0 & 0 & 0 & 0 & 0 & 0 & 0 & 0\\
1 & & 22 & 1658 & 334656 & 153601904 & 109900869120 & 
 119835268440320 & 179701691558215680 & 358810347942675191808\\
2 & & -6 & -2385 & -649384 & -348301926$\frac{2}{3}$
 & -276393269472 & -325054574102944 & -517144503241342976
 & -1083429067670151410841$\frac{3}{5}$\\
3 & & & 754 & 403500 & 284418460 & 266027954552  
 & $350188538012687\frac{1}{9}$ & 606338380910234368 
 & 1357922016436272064896\\
4 & & & -27 & -96524 & -108269551$\frac{2}{3}$ & -129814389840 &
 -200292896954500 & -388885779063580337$\frac{7}{9}$ 
 &-950703086097155610426$\frac{2}{3}$\\
5 & & & & 7860 & 20243976 & 35294435450 & 
 67835150491952$\frac{2}{3}$ & 153240568087470602$\frac{2}{3}$
 & 419124576286737422400\\
6 & & & & -108 & -1748436$\frac{2}{3}$ & -5461778850 
& -14237073600841 & -39235294734473786$\frac{2}{3}$
 & -123701080180932859946$\frac{2}{15}$\\
7 & & & & & 55980 & 465582548 & $1869074318751\frac{1}{3}$
 & 6694588679545816 & 25258034274069075192\\
8 & & & & & -405 & -19729020 & -150316130097
 & -764582679410893$\frac{1}{3}$ & -3623005587850488121\\
9 & & & & & & 326970 & $6996149842\frac{8}{9}$
 & 57616149115192 & 365799796385163936\\
10 & & & & & & -1458 & -168308595 & -2763401680710$\frac{2}{9}$
 & -25727820621001866$\frac{14}{15}$\\
11 & & & & & & & 1688526 & $78924402909\frac{1}{3}$ & 1230228773135952\\
12 & & & & & & & -5103 & -1196493480 & -38326543230336$\frac{2}{3}$\\
13 & & & & & & & & 8015112 & 725790362976\\
14 & & & & & & & & -17496 & -7483535892\\
15 & & & & & & & & & 35779320\\
16 & & & & & & & & & -59049\\
\br
\centre{10}{$\beta_{1,k,j}$} \\
\mr
0 & -1 & 0 & 0 & 0 & 0 & 0 & 0 & 0 & 0\\
1 & & 22 & 1874 & 380544 & 173844464 & 124703585280 & 
  135892573287680 & 203891746284011520 & 407060480146447669248\\
2 & & -6 & -2505 & -717448 & -387796806$\frac{2}{3}$
 & -309968480352 & -365371396244512 & -582626355682555904
 & -1221984137726510908569$\frac{3}{5}$\\
3 & & & 754 & 427116 & 309285900 & 293508255992 & $
388934364085487\frac{1}{9}$ & 676720632830112000 & 
1519953696088914094464\\
4 & & & -27 & -97964 & -114431071$\frac{2}{3}$
 & -140364933360 & -219200100776580 & -429068174740539313$\frac79$
 & -1054354166560941235770$\frac{2}{3}$\\
5 & & & & 7860 & 20801176 & 37316508410 & $72997883541272\frac{2}{3}$
 & $166848798285151882\frac{2}{3}$ & 459887247973724918464\\
6 & & & & -108 & -1759236$\frac{2}{3}$ & -5648008290 & -15047082876273 
 & -42102339808215418$\frac{2}{3}$ & -134133831039985827818$\frac{2}{15}$\\
7 & & & & & 55980 & 472540628 & $1940426816751\frac{1}{3}$
 & 7075210989955416 & 27042955126506979800\\
8 & & & & & -405 & -19793820 & -153559531937 &
 -795936397412877$\frac{1}{3}$ & -3828497810715795865\\
9 & & & & & & 326970 & $7058663482\frac{8}{9}$
 & 59143290605752 & 381544301457679104\\
10 & & & & & & -1458 & -168648795 & -2803243801350$\frac{2}{9}$
 & -26506146752612586$\frac{14}{15}$\\
11 & & & & & & & 1688526 & $79382817117\frac{1}{3}$ & 1253639579258096\\
12 & & & & & & & -5103 & -1198126440 & -38712986171808$\frac{2}{3}$\\
13 & & & & & & & & 8015112 & 728717117184\\
14 & & & & & & & & -17496 & -7490884212\\
15 & & & & & & & & & 35779320\\
16 & & & & & & & & & -59049\\
\br
\centre{10}{$\beta_{2,k,j}$} \\
\mr
0 & -1 & 8 & 0 & 0 & 0 & 0 & 0 & 0 & 0\\
1 & & 22 & 1826 & 399904 & 188173424 & 136763942784 & 
150135943911680 & 226399980054487040 & 453481872936027199488\\
2 & & -6 & -2553 & -747352 & -415492486$\frac{2}{3}$
 & -337028036672 & -400806757117984 & -643091153929042944
 & -1354435396148323267737$\frac{3}{5}$\\
3 & & & 754 & 438668 & 326540300 & 315331137752 &
 $422565255743151\frac{1}{9}$ & $741047725232481564\frac{4}{9}$
 & 1673542368950203659648\\
4 & & & -27 & -98972 & -118710111$\frac{2}{3}$ & -148615548440 
 & -235371247103300 & -465345164498565169$\frac{7}{9}$
 & -1151609375547099787322$\frac{2}{3}$\\
5 & & & & 7860 & 21201176 & 38883012026 & $77347956643752\frac{2}{3}$ 
 & $178970258611925237\frac{1}{3}$ & 497699046774849639872\\
6 & & & & -108 & -1767876$\frac{2}{3}$ & -5793139250 
& -15721316702945 & -44622273568721818$\frac{2}{3}$
 & -143694733990288250474$\frac{2}{15}$\\
7 & & & & & 55980 & 478155188 & $1999479461647\frac{1}{3}$
 & $7405891926422253\frac{1}{3}$ & 28659202377280586328\\
8 & & & & & -405 & -19848900 & -156259593217 &
 -822961941093677$\frac{1}{3}$ & -4012551877454570521\\
9 & & & & & & 326970 & $7112021962\frac{8}{9}$
 & $60457289130459\frac{5}{9}$ & 395524266443898240\\
10 & & & & & & -1458 & -168948171 & -2837772622950$\frac{2}{9}$
 & -27193733020204138$\frac{14}{15}$\\
11 & & & & & & & 1688526 & $79787198493\frac{1}{3}$ & 1274324332887664\\
12 & & & & & & & -5103 & -1199596104 & -39056845236000$\frac{2}{3}$\\
13 & & & & & & & & 8015112 & 731355327360\\
14 & & & & & & & & -17496 & -7497602676\\
15 & & & & & & & & & 35779320\\
16 & & & & & & & & & -59049\\
\br
\end{tabular}
\end{table}

This time we note that the 2 or 3 leading coefficients
are independent of $n$. 
And indeed we find that $\beta_{n,k,2k-j}/3^k$ 
is a polynomial in $k$ of order $2j+1$. In particular we have,

\[
\beta_{n,k,2k} = -3^k(k+1),
\]
\[
\beta_{n,k,2k-1} = 3^k(278k^3+99k^2-179k)/27,
\]
\[
\fl
\beta_{n,k,2k-2}  =  \left\{ \begin{array}{rl} 
  3^k(77284k^5-233636k^4+145247k^3+233636k^2-222531k)/1458  & n=0 \\
  3^k(77284k^5-233636k^4+148487k^3+233636k^2-225771k)/1458  & n=1 \\
  3^k(77284k^5-233636k^4+151727k^3+231692k^2-230955k)/1458  
    & n=2, \end{array} \right. 
\]
and
\[
\fl
\beta_{n,k,2k-3}  =  \left\{ \begin{array}{rl} 
3^k(107424760k^7-981604100k^6+3689847622k^5-5987330165k^4 & \\
+1103673490k^3+6968934265k^2-4900945872k)/590490 & n=0 \\
3^k(107424760k^7-981604100k^6+3703358422k^5-6035726045k^4 & \\
+1135273210k^3+7017330145k^2-4946056392k)/590490 & n=1 \\
3^k(107424760k^7-981604100k^6+3716869222k^5-6092228405k^4 & \\
+1180199050k^3+7073832505k^2-5004493032k)/590490
    & n=2. \end{array} \right. 
\]

So when calculating the extrapolation formulae \eref{eq:ef}
we first used the sequences for the correction terms to predict
as many polynomials as possible. When we ran out of terms we then
predicted as many of the leading coefficients from
\eref{eq:ap} and \eref{eq:bp} as possible. This in turn allowed us to find
more extrapolation formulae, which we could use (together with
the formulae for $\beta_{n,k,2k-j}$) to find more of the formulas for
the leading coefficients $a_{n,s,s+n+3-k}$ and
$b_{n,s,s+n-k}$. We repeat this until the process stopped with the 
extrapolation formulae listed above.

From $Z_{n,33}(u)$ we extended the series for $Z_0$ and $Z_1$
to ${\cal O}(u^{113})$,
while the series for $Z_2$ was extended to ${\cal O}(u^{112})$.
The resulting new low-temperature series terms are listed in 
table \ref{series}. The series terms for $n < 80$ can be found in
\cite{EGJ}. The full series is also available by electronic mail
or via the world wide web (see end of article for details).

\begin{table}
\caption{\label{series} New low-temperature series terms for
the square lattice spin-1 Ising magnetisation, $M(u)$, 
susceptibility, $\chi (u)$, and specific heat, $C_v (u)$.}
\tiny\rm
\begin{tabular}{lrrr}
\br
$n$ & $M(u)$ & $\chi (u)$ & $C_v (u)$ \\ 
\mr
80 & -54894921926791871723909
 & 1287288269903730631946751
 & 14980548594400026006446720 \\
81 & 126297750080690225982572
 & -2932618501022211032818300
 & -35701518428258787100072584 \\
82 & -92489409777625802742528
 & 2052947118516396940212072
 & 27948358015093163176475128 \\
83 & -292140800078967381434028
 & 7265393839008482331992336
 & 80787294138478683214133578 \\
84 & 1131329573810488998686811
 & -27533569403846138701663365
 & -328084218297769469375116200 \\
85 & -1662612103740713574604884
 & 39920035526199174994626036
 & 500413572618503616248896480 \\
86 & -492613608448080934983288
 & 14931369412552677008749932
 & 113883610899578636803467152 \\
87 & 8211147564410929129324192
 & -209438337886232625955934292
 & -2436189367728965310422489418 \\
88 & -18878709288285563929234997
 & 476343631011307763277303031
 & 5796339184388631113650158304 \\
89 & 13785042470417967980505100
 & -331962548955170787548255312
 & -4521670273968534171069307750 \\
90 & 43860793960742590039383898
 & -1182970662366618284903971030
 & -13149275656695239669180483520 \\
91 & -169598161959998919874255236
 & 4472491833451466220138614096
 & 53268752093378620751877315482 \\
92 & 249007028837325086293670283
 & -6471380268030636441237200097
 & -81105273455462683681912082728 \\
93 & 74990743692109664334064752
 & -2453067840623244506560874448
 & -18811719845845853329570528542 \\
94 & -1234426063083531162682558560
 & 34026585140776880977075866240
 & 395621948197278485674431575608 \\
95 & 2835867624597373150874747480
 & -77254798018355427358151075836
 & -939772638538221206856049448380 \\
96 & -2064573852622364936737424098
 & 53613918339192473420875219089
 & 730560439047179326970330734464 \\
97 & -6616142800429142888692342768
 & 192358286613090067676905433984
 & 2137403950614601807223987280454 \\
98 & 25541973318840653657077177270
 & -725486237294161236962098624258
 & -8637803751685796115480314923276 \\
99 & -37462784919716975353690569292
 & 1047757642093298413301695350332
 & 13128860187862258100931770296284 \\
100 & -11463726370314369427304114523
 & 402487643299242910269563033131
 & 3102521140136519939724246308400 \\
101 & 186381597774899493113385553664
 & -5520950047108379701377201734452
 & -64169713991043524782724303040306 \\
102 & -427804305737711686514509242390
 & 12514441143336335641931499176014
 & 152189962317529977423571879172796 \\
103 & 310497065688338308368382361752
 & -8648192214593796849228047528228
 & -117900563896358034274507714039430 \\
104 & 1002057377876891572217899261401
 & -31239827769954401229313918082438
 & -347042122337706823781434388021392 \\
105 & -3862092565977953294672165562608
 & 117548646091803611498084800222172
 & 1399138594951555677686779993658460 \\
106 & 5658355584526190756409048411218
 & -169446914107751509953285066975278
 & -2122963342745826385809572352025484 \\
107 & 1758697226344654211271247297588
 & -65942014127753928407511143000684
 & -510975035983101017637996662363534 \\
108 & -28247139203735773793168037850532
 & 894839314320840238617134121410478
 & 10397649426241825241936709101268192 \\
109 & 64775215576079739888972057107032
 & -2025063464948777359185732816409016
 & -24621560654630485557870861813977290 \\
110 & -46866364052133130252606723536110
 & 1393635387942882646442005194185966
 & 19008526767153717074391591041349580 \\
111 & -152306338289478778693447758360664
 & 5068325152947612595253210347568392
 & 56293290314061858138013078241465190 \\
112  & 586011194798187595036300267855234
 & -19027185558928704352441507167871531
 & -226417646512838638150966074051275808    \\
113 & -857572414389308013646509973160724
 &
 & 342972424908794790373332519783688916    \\
\br
\end{tabular}
\end{table}

\section{Analysis of the series}

We analysed the series using the same methods as in our
previous paper \cite{EGJ} to which we refer the reader for details. 
Here we will give only a short summary of the results including
improved estimates for the critical point and amplitudes.

The estimates $u_c=0.5540663(5)$ for the physical
singularity and $\beta = 0.12507(2)$ for the critical exponent of 
the spontaneous magnetisation were obtained from
homogeneous differential approximants (which are equivalent to
Dlog Pad\'e approximants) by averaging over $[N,M]$ approximants
with $|N-M| \leq 1$ using at least 100 series terms. The figure
in parenthesis represents the spread among the approximants
(basically one standard deviation) and should {\em not} be 
viewed as a measure of the true error as they cannot include
possible systematic sources of error. 
From these estimates it is clear
that $\beta = 1/8$ as expected. However, the estimates converge
very slowly towards this value and even with a series as long as
the present 114 terms the estimates have not yet settled down
to their true value and there is a slight downwards drift in
the estimates for both $u_c$ and $\beta$.
Analysis of the susceptibility and specific heat series yield 
exponent estimates fully in agreement with the expectations that 
$\gamma ' = 7/4$ and $\alpha ' = 0$. By using our knowledge of the
exact values of the critical exponents and assuming that close
to $u_c$ the estimates for the exponents depend linearly on
the estimates for the critical point (inspection of the various
approximants clearly supports this assumption) we are led to the
improved estimate for the critical point $u_c = 0.5540653(5)$.

We calculated the critical amplitudes using two different
methods, both of which are very simple and easy to implement.
In the first method, we note that if $f(u) \sim A(1-u/u_{c})^{-\lambda}$,
then it follows that
$(u_{c} -u)f^{1/\lambda}|_{u=u_{c}} \sim A^{1/\lambda}u_{c}$. So we
simply form the series for $g(u) = (u_{c} -u)f^{1/\lambda}$ and
evaluate Pad\'{e} approximants to this series at $u_{c}$. The result
is just $A^{1/\lambda}u_{c}$. This procedure works well for the
magnetisation and susceptibility series (it obviously cannot
be used to analyse the specific heat series) and yields the
estimates $A_M=1.20840(5)$ and $A_{\chi}=0.06172(4)$ where the
error bar primarily reflects the uncertainty due to the estimate 
of $u_c$. For the specific
heat series two different approaches have been used. In the
first approach we look
at the derivative of the specific heat series for which the
above method should work with $\lambda =1$. This yields 
the estimate $A_C=22.3(1)$.
In the second approach 
we start from $f(u) \sim A\ln (1-u/u_{c})$ and form the series
$g(u) = \exp (-f(u))$ which has a singularity at $u_c$ with 
exponent $A$. One virtue of this approach is that no prior estimate
of $u_c$ is needed. However, the spread among estimates from different
approximants is quite substantial, though the amplitude estimate is
consistent with that
listed above. Biasing the estimates at $u_c$ yields $A_C = 22.3(3)$.
In the second method, proposed by Liu and Fisher \cite{LF}, one
starts from $f(u) \sim A(u)(1-u/u_{c})^{-\lambda}+B(u)$ and then
forms the auxiliary function  $g(u) = (1-u/u_{c})^{\lambda}f(u)
\sim A(u) + B(u)(1-u/u_{c})^{\lambda}$. Thus the required amplitude
is now the {\em background} term in $g(u)$, which can be obtained
from inhomogeneous differential approximants \cite{serana}.
This method can also be used to study the specific heat series. One
now starts from $f(u) \sim A(u)\ln (1-u/u_{c})+B(u)$ and then looks
at the auxiliary function $g(u) = f(u)/\ln (1-u/u_{c})$. As before
the amplitude can be obtained as the background term in $g(u)$.
This analysis yields the amplitude estimates listed in
\tref{lfa}. With the exception of the specific heat amplitude
there is excellent agreement between these estimates and those
obtained from the first method. 

\Table{\label{lfa} Estimates for the physical critical amplitudes 
$A_M$, $A_{\chi}$ and $A_C$ from the method of Liu and Fisher obtained
from inhomogeneous first-order differential approximants.
L is the degree of the inhomogeneous polynomial.} 
\br
$L$ & \centre{1}{$A_M$} & \centre{1}{$A_{\chi}$} &\centre{1}{$A_C$} \\
\mr
5 & 1.208284(74) & 0.061712(47) & 20.20(26) \\
6 & 1.208270(15) & 0.06173(12) & 20.13(22) \\ 
7 & 1.208256(80) & 0.06170(20) & 20.23(23) \\ 
8 & 1.208266(16) & 0.06164(21) & 20.29(20) \\ 
9 & 1.20824(21) & 0.06170(12) & 20.28(32) \\  
10 & 1.208269(35) & 0.061866(60) & 20.31(26) \\  
15 & 1.208272(74) & 0.06165(15) & 20.47(32) \\
20 & 1.20824(12) & 0.06167(15) & 20.464(75) \\
25 & 1.208250(47) & 0.06172(11) & 20.448(46) \\  
30 & 1.208236(51) & 0.06174(20) & 20.42(12) \\
35 & 1.208228(60) & 0.061792(82) & 20.40(20) \\  
40 & 1.208263(70) & 0.06159(20) & 20.45(12) \\
\br
\endTable

Regarding the value of the confluent exponent $\Delta_1$ we have little
to add to our previous results. Even with a series as long
as 114 terms we could not obtain accurate estimates for $\Delta_1$.
Again the Baker-Hunter \cite{BH} transformed series of the magnetisation
favours a value around 1.05 while estimates from the suceptibility
series again fall in two groups around 1.15 and 1.4, respectively.
Using the transformation of Adler {\em et al} \cite{AMP}
\[
G(u) = \lambda F(u) + (u_c-u)\mbox{d}F(u)/\mbox{d}u,
\]
where $F(u)$ is the original series and $\lambda$ the leading critical
exponent, yields estimates consistent with $\Delta_1 =1$ for both
the magnetisation and susceptibility.

We find a non-physical singularity closer to the origin than $u_c$
at $u_{\pm} = -0.3019395(5)\pm 0.3787735(5)$ with exponents
$\beta = -0.1690(2)$, $\gamma ' = 1.1692(2)$ and $\alpha ' = 1.1693(3)$,
and a singularity on the negative $u$-axis at $u_- = -0.598550(5)$
with exponents equal to those at the physical critical point.
Note that our estimate for $\beta$ at $u_{\pm}$ has changed
substantially from that given in our previous paper. This is mainly 
because we have put greater emphasis on estimates obtained from
inhomogeneous first and second-order differential approximants.
We note that we now have firm evidence to show that the
scaling law $\alpha ' + 2\beta + \gamma ' = 2$ holds at both
the physical as well as the non-physical singularities.

\section*{E-mail or www retrieval of series}

The series for the spin-1 Ising model can be obtained via e-mail 
by sending a request to iwan@maths.mu.oz.au or via the world
wide web on http://www.maths.mu.oz.au/\~{ }iwan/ by following the
instructions.

\section*{Acknowledgements}

Financial support from the Australian Research Council is gratefully 
acknowledged.


\section*{References}

\begin{thebibliography}{99}

\bibitem{EGJ} Enting I G, Guttmann A J  and Jensen I 1994
\JPA {\bf 27} 6987

\bibitem{DNE} De Neef T and Enting I G 1977 \JPA {\bf 10} 801

\bibitem{BGDP} Baxter R J and Guttmann A J 1988 \JPA {\bf 21} 3193

\bibitem{E78} Enting I G 1978 \JPA {\bf 11} 563

\bibitem{serana} Guttmann A J 1989 {\em Phase Transitions and Critical
Phenomena} vol 13, ed C Domb and J Lebowitz (New York:Academic) pp 1-234

\bibitem{LF} Liu A J and Fisher M E 1989 {\em Physica} {\bf 156A} 35

\bibitem{BH} Baker G A and Hunter D L 1973 \PR B {\bf 7} 3377

\bibitem{AMP} Adler J, Moshe M and Privman V 1981 \JPA {\bf 17} 2233

\end{thebibliography}
\end{document}